\documentclass[pre,aps,twocolumn,longbibliography]{revtex4-1}

\usepackage[dvips]{graphicx}
\usepackage{amssymb,amsfonts,amsmath}
\usepackage{color}
\usepackage{ulem}
\usepackage[hidelinks]{hyperref}

\newcommand{\Jdot}{\dot{\textbf{J}}}
\newcommand{\J}{\textbf{J}}
\newcommand{\vel}{\textbf{v}}
\newcommand{\fext}{\textbf{f}_\text{ext}}
\newcommand{\fint}{\textbf{f}_\text{int}}
\newcommand{\divtau}{\nabla\cdot\boldsymbol{\tau}}

\begin{document}

\title{Custom Flow in Molecular Dynamics}

\author{Johannes Renner}
\affiliation{Theoretische Physik II, Physikalisches Institut,
  Universit{\"a}t Bayreuth, D-95440 Bayreuth, Germany}
\author{Matthias Schmidt}
\email{Matthias.Schmidt@uni-bayreuth.de}
\affiliation{Theoretische Physik II, Physikalisches Institut,
  Universit{\"a}t Bayreuth, D-95440 Bayreuth, Germany}
 \author{Daniel de las Heras}
\email{delasheras.daniel@gmail.com}
\homepage{www.danieldelasheras.com}
\affiliation{Theoretische Physik II, Physikalisches Institut,
  Universit{\"a}t Bayreuth, D-95440 Bayreuth, Germany}

\begin{abstract}
	Driving an inertial many-body system out of equilibrium generates 
	complex dynamics due to memory effects and the intricate relationships between the external driving
	force, internal forces, and transport effects. 
	Understanding the underlying physics is challenging and often requires carrying out case-by-case
	analysis. To systematically study the interplay between all types of forces that contribute
	to the dynamics, a method to generate prescribed flow patterns could be of great help.
	We develop a custom flow method to numerically construct the external force field
	required to obtain the desired time evolution of an inertial many-body system, as 
	prescribed by its one-body current and density profiles.
	We validate the custom flow method in a Newtonian system of purely repulsive particles
	by creating a slow motion dynamics of an out-of-equilibrium process and by 
	prescribing the full time evolution between two distinct equilibrium states.
	The method can also be used with thermostat algorithms to control the temperature.
\end{abstract}

\date{\today}

\maketitle

\section{Introduction}

The precise application of a space- and time-resolved external force 
field can be used to drive a many-body system out-of-equilibrium
in a controlled way. Analysing the response of a system to an external field is
a primary method to calculate transport coefficients~\cite{NNMD}
such as shear~\cite{hess2002determining,muller1999reversing} and bulk~\cite{PhysRevA.21.1756,doi:10.1063/1.4950760} viscosities, and the thermal conductivity~\cite{doi:10.1063/1.473271}.
Imposed pressure gradients, patterned substrates, capillary forces, electromagnetic fields,
and centrifugal forces are examples of external fields that can be used
in lab-on-a-chip devices~\cite{doi:10.1146/annurev.fluid.36.050802.122124}
for the control of microflows~\cite{RevModPhys.77.977}.
However, the dynamics are complex due to 
far from trivial relationships between the external driving,
the interparticle interactions, and transport
effects. It is therefore difficult to predict the time evolution of a many-body system under
the influence of an imposed external field, and case-by-case analyses
are often required~\cite{water1,PhysRevE.79.026305,doi:10.1063/1.4737927}.

We consider here the inverse problem: to impose
the desired dynamics and then find the corresponding external field.
Such inversion is a valuable tool even at the level of individual particles.
It allows for example the independent and simultaneous motion of several particles (that differ in either the
shape~\cite{Mirzaee-Kakhki2020} or the magnetic properties~\cite{Loehr2016}) in arbitrary directions using a single external field.
We focus here on many-body systems. We aim at specifying the dynamics of 
a many-body system, as given by the time evolution of the one-body density and the one-body current fields, 
and then find, with computer simulations, the corresponding space- and time-resolved external field.

From a fundamental view point, such inversion can be used to time-reverse the dynamics
of a many-body system~\cite{de2020flow} and might offer new insights on irreversible processes~\cite{toth2020exact,hoover2020time}.
The inversion can be also used as an alternative to Gauss's
principle of least constraint~\cite{Gauss} in order to impose constraints on a dynamical system.
From an applied view point, controlling the time evolution of the system instead of the external force acting on it
can also be useful for the calculation of transport coefficients
and relaxation times, especially
in nanochannels where deviations from the Navier-Stokes formalism and from
bulk behaviour are expected~\cite{G12,PhysRevLett.66.2758,doi:10.1063/1.458823,PhysRevE.61.1432,doi:10.1063/1.480758}.
The study of memory effects~\cite{lesnicki2016molecular,jung2017iterative}, the design of lab-on-a-chip devices~\cite{PhysRevLett.102.108304,li2010molecular},
and the determination of the slip length at the nanoscale~\cite{doi:10.1063/1.3675904} can also benefit from such inverse methodology.

From a theoretical perspective, the existence in equilibrium of a unique
mapping between the density distribution and the conservative external force 
forms the basis of quantum~\cite{HK1964,mermin1965thermal} and
classical~\cite{Evans1979} density functional theory.
In time-dependent quantum mechanical systems, the Runge–Gross
theorem~\cite{PhysRevLett.52.997} ensures the existence of a unique mapping
between the density distribution and a time-dependent external potential.
A classical analogue of the Runge-Gross theorem was proposed by Chan
and Finken~\cite{PhysRevLett.94.183001}.
The existence of a unique mapping between the kinematic fields
and the external force field plays a central role in power functional theory,
an exact variational principle for nonequilibrium classical many-body overdamped
Brownian~\cite{PowerF} and Hamiltonian systems~\cite{PFTMD} as well as for
many-body quantum systems~\cite{schmidt2015quantum}.

The rapid increase in computational power has made possible  
the development of numerical inverse methods that implement these unique mappings
in equilibrium for both classical~\cite{PhysRevLett.113.167801,de2019custom} and
quantum~\cite{PhysRevLett.100.153004,molecules24203660}
systems. It is hence possible 
to prescribe an equilibrium density distribution and find the corresponding
external potential that generates the density using e.g.\ Monte Carlo simulations in the case of
an equilibrium classical system.
We have also developed a custom flow method for time-dependent overdamped Brownian systems~\cite{de2019custom}.
The method is a valuable tool to generate specific flow and density patterns in a completely controlled
way~\cite{de2020flow}.
Custom flow is based on the exact one-body force balance equation that, in overdamped Brownian systems,
relates the friction (against the solvent) force field, the internal force field, the external force field, and
the thermal diffusion. Using Brownian dynamics simulations, custom flow finds the external force required to generate
the desired (imposed) time evolution of both the one-body density and the one-body current distributions.
We prescribe elements that enter into the force balance equation, namely the density and the current distributions, 
and use an iterative scheme to find the generating external force field.

The force sampling method~\cite{PhysRevLett.120.218001}
uses a closely related idea: by sampling the one-body internal force field and using the force balance equation it 
is possible to obtain the one-body density distribution of an equilibrium system. The density distribution
obtained using the forces acting on the particles~\cite{PhysRevLett.120.218001,doi:10.1080/00268976.2013.838316,borgis3d,Kofke,doi:10.1063/5.0029113}
are more accurate than those obtained via the traditional counting of particles at space points.

Here, we present a custom flow method for classical many-body
systems following Newtonian Dynamics. The method is motivated by the exact one-body
force balance equation, Sec.~\ref{sec:onebody}, and it constructs iteratively the
external force field that is required to generate the desired
(target) time-evolution of both the density and the current distributions, Sec.~\ref{CF}.
The method constitutes the solution of a complex inverse
problem in statistical physics and implements numerically
the map between the kinematic fields (density and current) and 
the external force field. Custom flow can be used with both conservative and 
non-conservative forces as well as with thermostats. 
We validate the method in a model system, Sec.~\ref{model}, of purely repulsive particles
using several test cases, Sec.~\ref{sec:Results}, including one with the Bussi-Donadio-Parrinello
thermostat~\cite{Parrinello}.

\section{Theory}\label{sec:theory}
\subsection{One-body force balance equation}\label{sec:onebody}
Consider a classical system with $N$ identical and mutually interacting particles
following Newtonian dynamics. The equations of motion of the $i-$th particle are
\begin{eqnarray}
	\frac{d\textbf{r}_i}{dt}&=&\frac{\textbf{p}_i}{m},\label{eq:N1}\\
	\frac{d\textbf{p}_i}{dt}&=&\textbf{f}_i,\label{eq:N2}
\end{eqnarray}
where $m$ is the mass of the particle, $\textbf{r}_i$ denotes its position,
$\textbf{p}_i=m\textbf{v}_i$ is the momentum of the particle with $\textbf{v}_i$ its
velocity, and $\textbf{f}_i$ is the total force acting on the particle,
\begin{equation}
	\textbf{f}_i=-\nabla_iu(\textbf{r}^N)+\fext(\textbf{r}_i,t),\label{eq:ftotal}
\end{equation}
which in general consists of an imposed time-dependent external contribution,
$\fext(\textbf{r}_i,t)$, and an internal contribution, $-\nabla_iu(\textbf{r}^N)$.
Here $\nabla_i$ is the partial derivative with respect to $\textbf{r}_i$ and
$u(\textbf{r}^N)$ is the interparticle potential energy with, $\textbf{r}^N=\{\textbf{r}_1\dots\textbf{r}_N\}$
the complete set of particle positions.

In molecular dynamics (MD) simulations the equations of motion,~\eqref{eq:N1} and~\eqref{eq:N2},
are integrated in time. The observables of interest can be obtained as space-
and time-resolved one-body fields. For example, the one-body density distribution is
given by
\begin{equation}
\rho(\textbf{r},t)=\left\langle\sum_{i=1}^N\delta\left(\textbf{r}-\textbf{r}_i\right)\right\rangle,\label{eq:density}
\end{equation}
where $\delta(\textbf{r})$ is the three dimensional Dirac delta distribution, the sum runs over all particles $N$, and
$\textbf{r}$ is the position vector. The brackets $\langle\cdot\rangle$ denote a statistical average, which out-of-equilibrium
is done at each time $t$ over different realizations of the initial conditions (that is, the positions and the velocities of the particles at
the initial time $t_0=0$).
Differentiation of eq.~\eqref{eq:density} with respect to time
yields the one-body continuity equation
\begin{equation}
\dot{\rho}(\textbf{r},t)=-\nabla\cdot\J(\textbf{r},t),\label{eq:MDcontinuity}
\end{equation}
where the overdot indicates a time derivative and the one-body current is defined as
\begin{equation}
\textbf{J}(\textbf{r},t)=\left\langle\sum_{i=1}^N\delta(\textbf{r}-\textbf{r}_i)\textbf{v}_i\right\rangle.\label{eq:current}
\end{equation}
Differentiating eq.~\eqref{eq:current} and using equations \eqref{eq:N1}, \eqref{eq:N2}, and \eqref{eq:ftotal} results
in the exact one-body force balance equation
\begin{align}
    m\Jdot(\textbf{r},t)=\rho(\textbf{r},t)\left[\fext(\textbf{r},t)+\fint(\textbf{r},t)\right]+\divtau(\textbf{r},t).\label{eq:MDonebodyforcebalance}
\end{align}
See e.g.\ Ref.~\cite{PFTMD} for a more detailed derivation of eq.~\eqref{eq:MDonebodyforcebalance}.
Here, the one-body internal force field is 
$\textbf{f}_{\text{int}}(\textbf{r},t)=\textbf{F}_{\text{int}}(\textbf{r},t)/\rho(\textbf{r},t)$,
where $\textbf{F}_{\text{int}}$ is the internal force density field given by
\begin{equation}
\textbf{F}_{\text{int}}(\textbf{r},t)=-\left\langle\sum_{i=1}^N\delta(\textbf{r}-\textbf{r}_i)
	\nabla_i u\left(\textbf{r}^N\right)
	\right\rangle.\label{eq:Fint}
\end{equation}
The last term in eq.~\eqref{eq:MDonebodyforcebalance} describes the transport effects that arise due to
the one-body description. This transport term involves the one-body kinetic stress tensor
\begin{equation}
	\boldsymbol{\tau}(\textbf{r},t)=-m\left\langle\sum_{i=1}^N\delta(\textbf{r}-\textbf{r}_i)\textbf{v}_i\textbf{v}_i\right\rangle,\label{eq:tau}
\end{equation}
where $\textbf{v}_i\textbf{v}_i$ is a dyadic product such that $\boldsymbol{\tau}$ is of second rank.

\subsection{Custom flow in inertial systems}\label{CF}
We present here an iterative method to construct the external force
that generates a prescribed time evolution of the one-body fields $\rho$ and $\J$.
The most general form of the iteration scheme reads
\begin{eqnarray}
	\fext^{(k+1)}(\textbf{r},t)&=&\fext^{(k)}(\textbf{r},t)+\alpha\left(\J(\textbf{r},t)-\J^{(k)}(\textbf{r},t)\right)\nonumber\\
	&&+\beta\left(\Jdot(\textbf{r},t)-\Jdot^{(k)}(\textbf{r},t)\right)\nonumber\\
	&&+\gamma\nabla\ln\frac{\rho(\textbf{r},t)}{\rho^{(k)}(\textbf{r},t)}.\label{eq:itegeneral}
\end{eqnarray}
Here $k$ denotes the iteration index  and $\alpha$, $\beta$, and $\gamma$ are free non-negative prefactors
that in general can carry spatial and temporal dependencies.
The fields $\rho$ and $\J$ are the prescribed (target) fields, and $\Jdot$ is also 
known since it follows directly from the prescribed $\J$ via partial time derivative.

The procedure to find $\fext(\textbf{r},t)$ iteratively using equation~\eqref{eq:itegeneral}
is the following: (i) Run an MD simulation, evolving all the initial microstates
from $t_0$ to $t_0+\Delta t$ and sampling 
the space-resolved one body fields $\rho^{(k)}(\textbf{r},t_0+\Delta t)$, $\J^{(k)}(\textbf{r},t_0+\Delta t)$,
and $\Jdot^{(k)}(\textbf{r},t_0+\Delta t)$; (ii) use the sampled fields at iteration
$k$ to construct the external force for the next iteration $k+1$ according
to eq.~\eqref{eq:itegeneral}, and (iii) iterate until the process converges and
the external force at time $t_0+\Delta t$ is found. Convergence is achieved once the
sampled fields $\rho^{(k)}$, $\J^{(k)}$, and $\Jdot^{(k)}$ are the same (within the
desired numerical accuracy) as their target counterparts $\rho$, $\J$, and $\Jdot$.
Next, advance the time from $t_0+\Delta t$ to $t_0+2\Delta t$ 
and repeat the previous steps until the external force at $t_0+2\Delta t$ is found.
The process is repeated for the complete time evolution which is discretized in
time steps $\Delta t$. 

The idea behind Eq.~\eqref{eq:itegeneral} is simple but very useful: at each time,
the external force at iteration $k+1$ is that at the previous iteration $k$ plus
terms that (i) correct the deviations in the sampled fields with respect to the target
fields and (ii) vanish if the target and sampled fields are identical.
For example, if the current at a given position is higher (lower) than
the desired one, the external force at that position decreases (increases) in the next iteration.
Other correction terms are possible provided that they change the external force
at each iteration in the right direction. For example, the third term on the right
hand side of eq.~\eqref{eq:itegeneral} can be replaced by something like $\gamma\nabla(\rho-\rho^{(k)})$.
The precise form of eq.~\eqref{eq:itegeneral} is motivated by the exact force
balance equation, as we show in the Appendix~\ref{append}.

The non-negative prefactors $\alpha$, $\beta$, and $\gamma$ control how much
the external force changes in one iteration.
Using the exact one-body force balance equation~\eqref{eq:MDonebodyforcebalance} we
obtain suitable expressions for them
(see the Appendix~\ref{append} for a detailed calculation):
\begin{eqnarray}
	\alpha(\textbf{r},t) &=& \frac m{\rho(\textbf{r},t)\Delta t},\label{eq:alpha}\\
	\beta(\textbf{r},t) &=& \frac{m}{\rho(\textbf{r},t)},\\
	\gamma &=& k_BT_0.
\end{eqnarray}
Here $k_B$ is the Boltzmann constant and $T_0$ denotes the temperature of 
the initial state. Recall that the above expressions for $\alpha$, $\beta$, and
$\gamma$ are not unique. The prefactors only fix the amount of change
between two iterations. The method can in principle be also implemented by
simply using non-negative constant prefactors. Also, not all three prefactors
need to be present. Actually, having only $\alpha$ or only $\beta$ is 
sufficient to find the external force iteratively. In cases for which the
time dependent density distribution $\rho$ determines the full dynamical evolution
of the system it is also possible to work only with the coefficient $\gamma$. Such 
cases occur only if the current is free of both rotational and harmonic terms such
that $\rho$ alone fully determines $\J$ via eq.~\eqref{eq:MDcontinuity}.

In our particular implementation of Eq.~\eqref{eq:itegeneral} we iterate using only the target
and sample currents. Hence, we set $\alpha$ to the value in Eq.~\eqref{eq:alpha} and set 
both $\beta$ and $\gamma$ to zero. Then, the iterative custom flow method we use here reads
\begin{equation}
    \fext^{(k+1)}(\textbf{r},t)=\fext^{(k)}(\textbf{r},t)+\frac{m}{\rho(\textbf{r},t)\Delta t}\left(\J(\textbf{r},t)-\J^{(k)}(\textbf{r},t)\right),\label{eq:fextJ}
\end{equation}
This iteration scheme is repeated at every $\Delta t$.
That is, we use eq.~\eqref{eq:fextJ} to iteratively find $\fext(\textbf{r},t_0+\Delta t)$. We then advance
time to $t_0+2\Delta t$ and use eq.~\eqref{eq:fextJ} to find $\fext(\textbf{r},t_0+2\Delta t)$. The process
repeats until the complete time evolution is found.

The same algorithm can be used to find a suitable collection
of initial microstates at $t_0$ such as e.g. microstates from an equilibrium system
with a prescribed one-body density distribution.
To this end we can start with an homogeneous equilibrium
system and use custom flow to find microstates of another equilibrium system
with the  desired density distribution that serves as our initial state at $t_0$.
Alternatively, such initial set of microstates can be also found using
the inversion between the external field and the density
distribution for equilibrium systems described in Ref.~\cite{de2019custom}.
 
At each time we initialize the iterative process ($k=0$) using the external force
\begin{equation}
\fext^{(0)}(\textbf{r},t)=\frac{m\Jdot(\textbf{r},t)}{\rho(\textbf{r},t)},\label{eq:fextinitial}
\end{equation}
which follows by making both $\fint$ and $\divtau $ zero everywhere
in eq.~\eqref{eq:MDonebodyforcebalance}.
Another possible initialization is to
include the contributions of the internal force and the kinetic stress tensor
at the previous sampling time step $\Delta t$ on the right hand side of eq.~(\ref{eq:fextinitial}).

In principle the time step $\Delta t$ can be as small
as the integration time step $dt$ of the MD simulation. In practice,
however, we use a larger time step ($\Delta t/ dt=10$) that does not compromise
the accuracy of the calculation but still reduces the computational effort.
In the MD integration algorithm of the equations of motion, we keep the external force
constant between two consecutive time steps, $t$ and $t+\Delta t$.
Interpolating the external force between two consecutive time steps
can be problematic in cases where $\fext$ changes drastically (such as 
e.g. if an external force is switched on at a specific time).

The iteration scheme \eqref{eq:fextJ} is particularly well suited since
it is general and it requires to sample only
$\J^{(k)}$ in order to find the external
force for the next iteration $k+1$. Sampling of
$\Jdot$, that can be done by individually sampling the terms on
the right hand side of eq.~\eqref{eq:MDonebodyforcebalance}, is therefore not required.
The proposed version of custom flow in eq.~\eqref{eq:fextJ}
does not require knowledge of any additional contribution
or modification to the one-body force balance equation that might arise
if e.g.\ a thermostat algorithm acts on the many-body level. As we demonstrate
below our method works also with thermostats.

\section{Model and simulation details}\label{model}

We implement the method in a system of $N=50$ particles that interact via the purely repulsive Weeks-Chandler-Anderson
interparticle-interaction potential~\cite{WCA-potential}:
\begin{align}
    \phi(r)= 
\begin{cases}
	4\epsilon\left[\left(\frac{\sigma}{r}\right)^{12}-\left(\frac{\sigma}{r}\right)^6+\frac14\right],& \text{if } r\leq r_c\\
    0,&              \text{otherwise}.
\end{cases}
\end{align}
Here, $r$ is the distance between two particles,
$\sigma$ is the length scale, $\epsilon$ is the energy scale, and $r_c/\sigma=2^{1/6}$ is the
cutoff radius which is set at the minimum of the 12-6 Lennard-Jones potential.
We use $\tau=\sqrt{m\sigma^2/\epsilon}$ as the time scale.

The particles are located in a periodic three-dimensional
simulation box with lengths $L_x$, $L_y$, and $L_z$. The origin of coordinates
is located at the center of the simulation box. The system is
inhomogeneous only in the $\mathbf{\hat x}$-direction, which is discretized
with bins of size $0.05\;\sigma$. The system remains homogeneous 
and without average flow in the other two directions, $\mathbf{\hat y}$ and $\mathbf{\hat z}$.

The many-body equations of motion, eqs.~\eqref{eq:N1} and~\eqref{eq:N2}, are
integrated in time using the velocity Verlet algorithm~\cite{VelVerlet} with an integration
time step $dt/\tau={10^{-4}}$.
The particle positions
are initialized randomly with the condition that the particles do not interact with each
other (all interparticle distances are larger than $r_c$). The particle velocities
are initialized according to a Maxwell-Boltzmann distribution. Hence, each velocity
component is generated from a Gaussian distribution with zero mean and
standard deviation $\sqrt{2k_BT/m}$ with $T$ absolute temperature, which needs to be prescribed.
The center of mass is set initially at rest.
We then let the system equilibrate for $1\,\tau$ with no external force applied.

For the custom flow method we use a time step of $\Delta t/dt = 10$
and average at each time over $2\cdot10^6$ trajectories
(different realizations of the initial positions and velocities of the particles at $t=0$)
to obtain accurate results.
Since the time step $\Delta t$ is small and the prefactor $\alpha$ has
been carefully selected, only three iterations are required at each time
for the method to converge to the desired external force.
The one-body fields are sampled at every $\Delta t$ and used according to
eq.~\eqref{eq:fextJ} to find the external force for the next iteration. 

\section{Results}\label{sec:Results}
We illustrate the validity of the method with three examples.
In Sec.~\ref{IIIA} we measure the time evolution of the one-body fields $\rho$
and $\J$ in a system subject to a spatially inhomogeneous external force which is switched on at $t=0$ and then kept constant in time.
We next construct the time-dependent external force required to slow down the
observed dynamics by an arbitrary factor. 
In Sec.~\ref{IIIC}, we incorporate a thermostat to demonstrate its easy implementation
within custom flow. Finally, in Sec.~\ref{IIID}, we prescribe the full time evolution of the one-body
density and the one-body current and then find the corresponding external force.

\subsection{Slow motion dynamics}\label{IIIA}

\begin{figure*}
    \centering
    \resizebox{\textwidth}{!}{\includegraphics{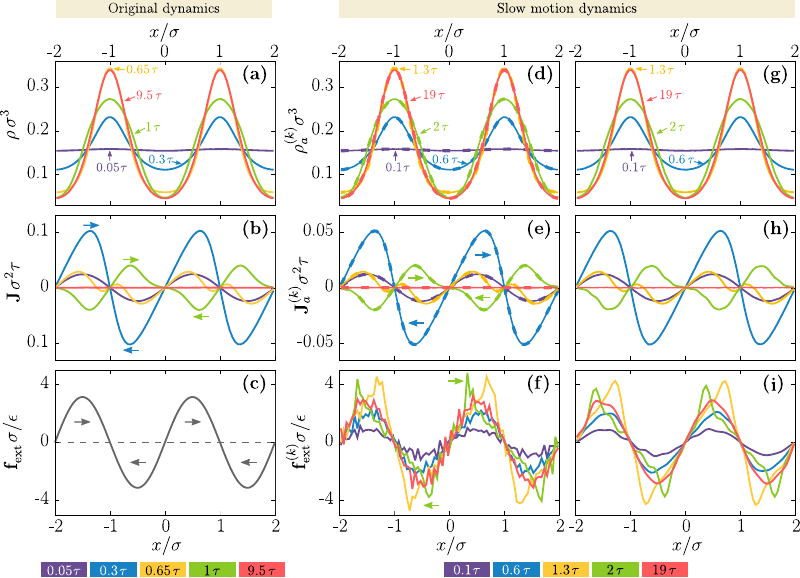}}
	\caption{Time evolution of the density (a) and the current (b) profiles
	of a system under the influence of a sinusoidal external force (c).
	The profiles in (a) and (b) are colored according to time, as indicated.
	Both $\J$ and $\fext$ act along the $\mathbf{\hat x}$-axis.
	The horizontal arrows illustrate the direction of the vector field
	at specific regions (position of the arrow) and also times (color).
	Sampled (solid-lines) density (d) and current (e) profiles of a slow motion system with
	dynamics slowed down by a factor $a=1/2$ (note that the indicated times
	are twice those in panels (a) and (b)). The dashed lines indicate the target profiles.
	(f) External force, color-coded as in (d), obtained with custom flow,
	that produces the slow motion dynamics. 
	Sampled density (g) and current profiles (h) for a system
	under the influence of the external force field (i)
	which is a smoothed version of that shown in (f).
	The set of initial microstates used to obtain the averaged fields
	in the second and in the third column are not the same. A video showing the
	time evolution is provided in the Supplemental Material~\cite{Supp}.}
    \label{fig1}
\end{figure*}

As a first example, we use custom flow to modify the 
time scale of a dynamical process.
The particles are located in a box with dimensions
$L_x/\sigma=4$, $L_y/\sigma=8$, and $L_z/\sigma=10$. 
We start with a homogeneous
system at equilibrium at $t=0$ and initial temperature $k_BT/\epsilon=0.5$.
Then, we switch on the following
external potential
\begin{equation}
    V(x)=V_0\cos\left(\frac{4\pi x}{L_x}\right),
\end{equation}
with $V_0/\epsilon=1$. Hence, the corresponding external force, which is constant
in time for $t>0$, acts only in the $\mathbf{\hat x}$-direction
\begin{equation}
	\fext(x)=-\nabla V(x)=\frac{4\pi}{L_x}\sin\left(\frac{4\pi x}{L_x}\right)\mathbf{\hat x}.\label{eq:originalfext}
\end{equation}
We let the system evolve for a total time of $10\,\tau$,
which is long enough to reach proximity
to a new equilibrium state with an inhomogeneous density profile.
The time evolution of the one-body density and current profiles are shown
in Fig. \ref{fig1}a) and~\ref{fig1}b), respectively.
The external force field, shown in Fig.~\ref{fig1}c),
accelerates the particles toward the minima of the external potential, located at
$x/\sigma=\pm 1$. 
Two density peaks grow from the homogeneous density distribution at $t=0$, reaching
the largest amplitude at  $t/\tau\approx0.65$. At short times, the shape of the one-body current
resembles that of the external force (sine wave) and it increases in amplitude until it reaches its largest value
at $t/\tau\approx0.3$. Then, the amplitude of the current decreases
until the current starts to flip sign, which occurs when the density peaks reach the largest amplitude ($t/\tau\approx0.65$).
Next, the density peaks decrease in amplitude and get broader since only some particles have enough
momenta to overcome the external potential barrier.
Most particles, however, cannot overcome the potential barrier. Instead, they climb
partially the barrier (contributing to the broadening of the density peaks). Then,
once the kinetic energies of the particles have been transformed into external energy,
the particles start to move backwards towards the minima of the potential. This backward
motion leads to an increase of the density peaks and the process repeats again in time.
Now, however, the process is less intense since both the energy stored
in the current and the external energy have been partially dissipated due to e.g. interparticle collisions
and converted into thermal energy. The described time evolution repeats in time creating a
damped oscillatory behaviour. 
Eventually, the system reaches an equilibrium state at $t/\tau\approx9.5$ with vanishing
one-body current (within our numerical accuracy). Recall that this highly nontrivial time evolution of the one-body density and current
profiles is produced by a simple external force, see eq.~\eqref{eq:originalfext}, that is switched on at $t=0$ and then kept constant in time.
A video showing the time evolution of the one-body fields is provided
in the Supplemental Material~\cite{Supp}.

Next, we use custom flow to find the external force
required to reproduce this complex dynamics but in slow motion, i.e. slowed down by an arbitrarily
chosen factor. By changing the time scale of the process we expect the external force
of the slow motion system to be time dependent, which is indeed the case. 
We want to scale the time by a factor $a$. Hence, in the
new system the density profile $\rho_a$ at time $t$ is the same as 
the density profile in the original system at time $at$. That is,
$\rho_a(\textbf{r},t)=\rho(\textbf{r},at)$. The time derivative in the continuity equation~\eqref{eq:MDcontinuity}
implies that the current in the new system is also scaled by a factor $a$. That is,
$\J_a(\textbf{r},t)=a\J(\textbf{r},at)$. Therefore,
scaling the time leads to a factor $a$ in front of the current that 
needs to be considered when prescribing the target fields.
Similarly, the time derivative 
of the current gets an additional scaling factor $\Jdot_a(\textbf{r},t)=a^2\Jdot(\textbf{r},at)$.

In Figs.~\ref{fig1}d) to \ref{fig1}f) 
we show the slow motion dynamics with scaling factor $a=0.5$. Hence, the slow motion
runs for $20\;\tau$, i.e. twice the original total time.
The sampled one-body density and current profiles coincide with their target.
The time evolution of the one-body density, Fig.~\ref{fig1}d), is the same as in the
original system, Fig.~\ref{fig1}a), but it proceeds only at half speed.
Similarly the evolution of the one-body current, Fig.~\ref{fig1}e)
is two times slower and the amplitude is half of the original target current profiles, Fig.~\ref{fig1}d).
The time dependent external force that generates the slow motion
(found with custom flow) is shown in Fig.~\ref{fig1}f).
At short times the shape of the external force resembles that in the original system, Fig.~\ref{fig1}c), 
but its amplitude is reduced by a factor four. 
This was expected since at $t=0$ the system is in equilibrium 
under no external force. That is, at $t=0$ both $\divtau$ and $\textbf{F}_{\text{int}}$ are 
homogeneous and cancel each other
in eq.~\eqref{eq:MDonebodyforcebalance}. At short times, only
$\Jdot_a$ contributes to the external force in the force balance
equation and hence the external force has the same shape as in the original
system but it is rescaled by a factor $a^2$ since $\Jdot_a=a^2\Jdot$ as discussed above.

Although the maximum amplitude of the density peaks occurs at $1.3\,\tau$,
the amplitude of the external force continues to grow and its shape deviates
from a sinusoidal wave. The maxima and the minima of the external force are
shifted towards the location of the density peaks at $x/\sigma=\pm 1$.
While the amplitude of the density profile decreases,
the extrema of the external force shift towards the minima
of the one-body density at $x=0$ and $x/\sigma=\pm 2$.

When the slow motion system reaches the equilibrium state,
the shape of the external force resembles that in the original
system but, interestingly, the amplitude is slightly
smaller in the slow motion system. In slow motion 
less energy is dissipated due to the reduced value of
the one-body current. Hence, also the temperature is slightly different.
For the original time evolution the final temperature after equilibrium is reached
is $k_BT/\epsilon=0.77$ and in slow motion it is $k_BT/\epsilon=0.68$ ~\footnote{We calculate the temperature in equilibrium via the equipartition theorem
$E_{\rm kin}=\frac32(N-1)k_BT$, where $E_{\rm kin}$ is the total kinetic energy of the system
$E_{\rm kin}=\left\langle\frac m2\sum_{i=1}^N\vel_i^2\right\rangle$.
}. This temperature difference 
is responsible for the different amplitudes of the 
external force. Note that in equilibrium the transport term reduces to
\begin{equation}
	\divtau=-k_BT\nabla\rho.\label{eq:divtau}
\end{equation}
Hence, according to eq.~\eqref{eq:MDonebodyforcebalance}
it is clear that two equilibrium systems with the 
same density distribution but at different temperatures
are generated by external forces with different amplitudes.

{\bf Robustness of external force.}
Custom flow generates a noisy external force, see Fig.~\ref{fig1}f).
The iterative scheme, eq.~\eqref{eq:fextJ}, minimizes the error in the current profile
since it is designed to converge if target and sampled currents coincide.
As a result, the statistical fluctuations in $\J$ are directly translated
into the external force field. Such fluctuations arise due to the finite
number of realizations (recall that at each time we average over $2\cdot10^6$ trajectories
that have evolved from different realizations of the initial particle positions
and velocities).
One could then think that the external force is tailored to the
set of initial conditions and cannot be used in other circumstances.
We demonstrate in the following that this is not the case and 
that the external force is indeed quite robust and independent
of the small details.
To this end we first smooth the external force in Fig.~\ref{fig1}f)
using a Fourier transform and eliminating the high frequency modes
(only the lowest $15$ modes are retained). The smoothed external force is depicted
in Fig. \ref{fig1}i).
Next, using the smoothed external force and also a different set 
of initial conditions (with the same number of microstates, i.e. $2\cdot 10^6$) we sample
the time evolution of the density and the current one-body fields, shown
in Figs.~\ref{fig1}g) and \ref{fig1}h), respectively.

Both the density and the current profiles obtained with the original external force and with the
smoothed external force and a different set of initial conditions are similar and reproduce accurately
the target profiles. Of course small differences occur, compare e.g.\ the current
profiles at $t/\tau=1$ in Figs.~\ref{fig1}e) and~\ref{fig1}h).
Hence, we conclude custom flow is suitable
to reproduce the prescribed dynamics using other sets of initial
conditions provided that enough trajectories are used.
Furthermore, we want to stress that for the cases considered here
the noise in the external force
is not relevant to generate the target fields $\rho$ and $\J$. We
expect that other filters that keep the structure and eliminate
the noise can also be used to smooth the external force.

\subsection{Thermostats}\label{IIIC}
It is often the case that MD simulations are performed at
constant temperature. In the following we show custom flow is also valid with algorithms
to control the temperature (thermostats).
Several types of thermostats can be implemented in MD~\cite{thermostattemperature}.
In general, a thermostat acts on the many-body level by rescaling the particle velocities and modifying
therefore the equations of motion and the integration algorithm. Hence, thermostats generate
new terms in the one-body force balance equation~\eqref{eq:MDonebodyforcebalance}.
However, custom flow is designed such that sampling of these terms is not required to 
advance the iterative process.
Custom flow uses only
the external force at the previous iteration and the sampled current field, cf. eq.~\eqref{eq:fextJ}.
The external force constructed with custom flow changes, of course, if a thermostat is used but
the implementation of the method remains unchanged.

To illustrate the use of thermostats, we implement the well known Bussi-Donadio-Parrinello (BDP) thermostat~\cite{Parrinello},
an extension of the Berendsen thermostat~\cite{Berendsen}, that stochastically ensures a thermalized distribution of the kinetic energy.

Out of equilibrium, the kinetic energy has a contribution due to the net flow of the system
\begin{equation}
    E_{\text{flow}}=\frac{m}{2}\int d\textbf{r}\rho\vel^2,
\end{equation}
which is unrelated to the temperature~\cite{thermostattemperature}. To control
the temperature considering only the velocity fluctuations around
the mean velocity~\cite{loose1992temperature}, we can
rescale the particle velocities using only the thermal kinetic energy:
\begin{equation}
    E_{\text{thermal}}=\left\langle\frac{m}{2}\sum\limits_{i=1}^N (\vel_i-\vel)^2\right\rangle,\label{eq:Ekinthermal}
\end{equation}
where it is important to note that $\vel(\textbf{r},t)$ is the space- and time-dependent
velocity profile (and not the center of mass velocity).
The implementation of eq.~\eqref{eq:Ekinthermal} is particularly
simple within custom flow since the velocity profile is known in advance.

\begin{figure*}
    \centering
    \resizebox{\textwidth}{!}{\includegraphics{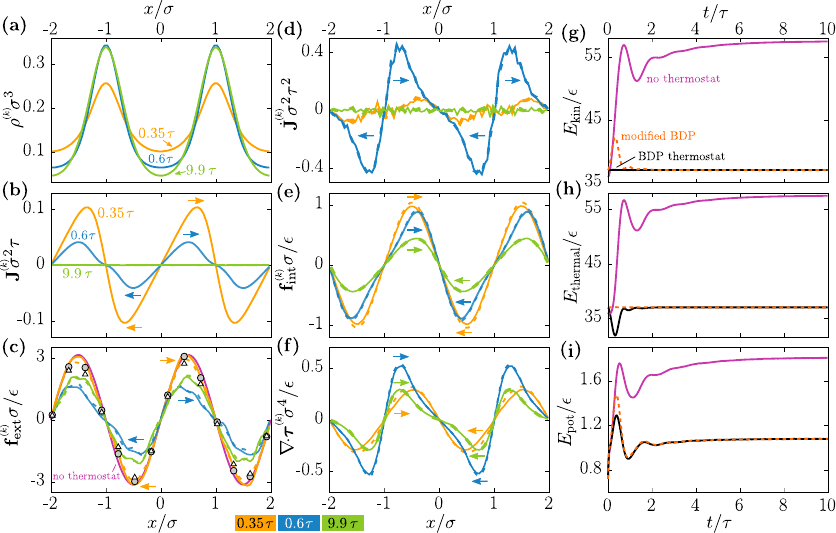}}
	\caption{Time evolution of a system with target $\rho$ and $\J$ as those in panels (a) and (b) of Fig.~\ref{fig1}, respectively, but
	under the influence of either the BDP thermostat (solid lines) or the modified BDP thermostat (dashed lines).
	Three different times are displayed: $t/\tau=0.35$ (yellow), $t/\tau=0.6$ (blue), and $t/\tau=9.9$ (green).
	(a) density profile $\rho$, (b) current profile $\J$, (c) external force $\fext$, (d) time derivative of the current $\Jdot$,
	(e) internal force $\fint$, and (f) transport term $\divtau$. All vector fields act along the $\mathbf{\hat x}$-axis.
	The horizontal arrows in panels (c) to (f) indicate the direction of the respective field at specific regions and times,
	as indicated by the position and the color of the arrow, respectively.
	The violet line in (c) indicates the constant in time external force that generates the target fields $\rho$ and $\J$ if 
	no thermostat is present. The grey circles (white triangles) in (c) show the external force calculated via the force balance
	equation for the BDP (modified BDP) thermostat at time $t/\tau=0.35$.
	The third column shows the time evolution of the kinetic (g), thermal (h), and potential (i) energies
	in the case of no thermostat (violet), BDP (black) and modified BDP (dashed orange) thermostats.
	}
    \label{fig2}
\end{figure*}

To demonstrate that custom flow can be used with thermostats,
we find the external force that generates the same time
evolution of $\rho$ and $\J$ as that in Figs.~\ref{fig1}a) and ~\ref{fig1}b)
but using the BDP thermostat.
We set the time constant (required in the algorithm to control the temperature) to five times
the integration time step $dt$ of the simulation.
We show in Fig.~\ref{fig2} results from both using the total kinetic energy and only the
thermal energy, eq.~\eqref{eq:Ekinthermal}, to rescale the velocities.

The sampled one-body density and one-body current are
displayed in Figs.~\ref{fig2}a) and \ref{fig2}b), respectively.
The results from both thermostats are on top of each other
and are indistinguishable from the case without a thermostat, Fig.~\ref{fig1}.
In Fig.~\ref{fig2}c) we show the external force constructed with custom flow
and smoothed using the same procedure as discussed above. We have verified that
the smoothed external force also reproduces the target time evolution.
The inclusion of the thermostat has a strong influence on the external force,
which is now time dependent, in contrast to the original (constant energy)
dynamics for which the external force is constant in time, eq.~\eqref{eq:originalfext}.
Furthermore, the external force required to produce the desired dynamics depends on
the thermostat.
This is clearly visible at e.g.\ $t/\tau=0.35$
at which the current profile has its largest amplitude.
At that time the flow kinetic energy has also its largest value and hence
the two versions of the thermostat rescale the velocities differently.
Differences are also noticeable in both the internal force field, see Fig.~\ref{fig2}e), and
especially in the transport term $\divtau$, Fig.~\ref{fig2}f),
since ${\boldsymbol{\tau}}$ is directly related to the total kinetic energy
cf. eq.~\eqref{eq:tau}.

Custom flow can help to understand how different thermostat algorithms modify
the physical properties of a system. In Fig.~\ref{fig2}c) we plot
the external force at $t/\tau=0.35$ calculated by using the sampled
$\fint$ and $\divtau$ into the force balance
equation \eqref{solvefext}, which is exact only
without thermostats. Using the thermal kinetic energy in the BDP thermostat
results in an approximated external force, via eq.~\eqref{solvefext}, that is almost on top of the actual force
generated with custom flow. The external force obtained with the original BDP thermostat (that uses the total kinetic energy)
via the force balance equation~\eqref{solvefext} shows a clear deviation from the force obtained in custom flow.
We therefore conclude that using the total kinetic energy in the BDP
thermostat induces a nontrivial contribution to the force balance equation
that alters the flow.
In contrast, using the thermal energy, only the velocity fluctuations are rescaled
and the flow is left unchanged.

As expected, the time derivative of the current $\Jdot$, Fig.~\ref{fig2}d), is very 
small for  $t=0.35\;\tau$ since $\J$ reaches at that time 
the maximum amplitude. Also for $t/\tau\ge9.9$ 
the system is very close to equilibrium and $\Jdot$ vanishes within
the numerical accuracy.

Finally, we show in Figs.~\ref{fig2}g), ~\ref{fig2}h), and ~\ref{fig2}i)
the time evolution of the total kinetic energy, the thermal energy, and the
internal potential energy, respectively. Shown are the original dynamics (no thermostat)
and both the BDP thermostat and the modified version that uses only the thermal kinetic energy.
The total kinetic energy is constant in time for the BDP thermostat as it should be
by construction. In the modified version, the thermal kinetic energy, Fig.~\ref{fig2}g), is constant in time
but the total kinetic energy varies with time since the flow kinetic energy is kept unchanged.
The original time evolution is clearly not at constant temperature since both
the kinetic and the thermal kinetic energy vary substantially over time. The total 
internal potential energy is for neither thermostat constant in time, Fig.~\ref{fig2}i).
For a short period of time around $t/\tau=0.35$ there is a
significant difference between both versions of the BDP thermostat due to the large amplitude
of the one-body current. 

Custom flow can be used as a new tool to analyse the quality and the
physical consequences of the inclusion of thermostats in the dynamics of many-body
systems. We have shown here that separating the flow and
thermal kinetic energies, especially in systems with large magnitude
of the one-body current, is advisable.

\subsection{Tailoring inhomogeneous density profiles}\label{IIID}
In the previous examples we obtained $\rho$ and $\J$ in a
simulation for a fixed external force and used modified versions of
them as target fields. In this last example, we show there is 
freedom to prescribe the fields provided
that they represent a physical system. For example, the target $\rho$ and
$\J$ must obey the continuity equation.

We set a simulation box with dimensions $L_x/\sigma=10$, $L_y/\sigma=5$ and $L_z/\sigma=5$,
and prescribe the one-body density 
\begin{align}
	\rho(x,t) =\rho_0 -\frac{A}{2}\cos\left(\frac{4\pi x}{L_x}\right)\left[1-\cos\left(\frac{\pi t}{T_0}\right)\right],\label{eq:rhoevol}
\end{align}
with average density $\rho_0=N/(L_xL_yL_z)=0.2\,\sigma^{-3}$, and
constants $A\sigma^3=0.05$ (maximum amplitude of the density inhomogeneity)
and $T_0/\tau=0.5$. At $t=0$ the density is homogeneous, see eq.~\eqref{eq:rhoevol}, and the system
is in equilibrium. The density profile evolves according to eq.~\eqref{eq:rhoevol}
for $0<t<T_0$ (two peaks grow from the initial homogeneous state).
At $t\ge T_0$ the one-body current is set to zero everywhere and
therefore the inhomogeneous density profile remains stationary.

The target current $\J$ follows from eq.~\eqref{eq:rhoevol} and
the space integral of the continuity equation~\eqref{eq:MDcontinuity}: 
\begin{align}
	J_x=-\int dx\;\dot{\rho} + C,\label{eq:currenttarget}
\end{align}
with $J_x$ the $x$-component of $\J$ and $C$ a constant that we set such
that the total integral of the current vanishes $\int dxJ_x=0$.
That is, for convenience we choose to not have motion of the center of mass.
We calculate the target current analytically using eq.~\eqref{eq:currenttarget}.
Note that in our effective one-dimensional system with periodic boundary conditions
the time evolution of
$\rho$ determines the current $\J$ up to a constant only. However, in
higher dimensions the continuity equation alone is not enough to determine
the current from the time-evolution of the density profile since any
divergence-free field can be added to the current without altering
$\rho$.

\begin{figure*}
    \centering
	\resizebox{\textwidth}{!}{\includegraphics{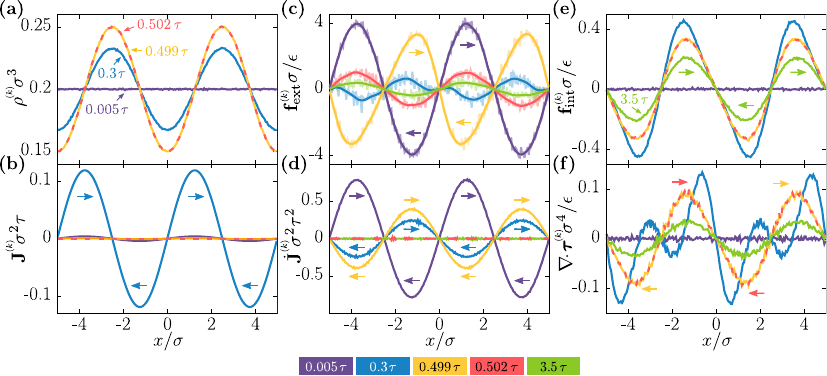}}
	\caption{Sampled one-body (a) density $\rho$ and (b) current $\J$ profiles at different
	times, as indicated. The target fields are prescribed according to eqs.~\eqref{eq:rhoevol} and~\eqref{eq:currenttarget}.
	(c) External force field constructed with custom flow (semitransparent lines) and smoothed external force (solid lines).
	Sampled (d) time derivative of the current profile $\Jdot$, (e) internal force
	density $\fint$, and (f) transport term $\divtau$. All vector fields act along
	the $\mathbf{\hat x}-$axis. The horizontal arrows indicate the direction of the respective
	field at specific regions (arrow's position) and times (arrow's color). The Supplemental Material~\cite{Supp} contains
	a movie showing the time evolution.}
    \label{fig3}
\end{figure*}

Using custom flow to construct
the external force that generates the
time evolution prescribed in eq.~\eqref{eq:rhoevol}
yields the results shown in Fig. \ref{fig3} (a movie is also
included in the Supplemental Material~\cite{Supp}).
Panels a) and b) of Fig.~\ref{fig3} show for different times
the sampled one-body density and the one-body current, respectively.
Both fields are in perfect agreement with their respective target fields,
cf. eqs.~\eqref{eq:rhoevol} and~\eqref{eq:currenttarget}.
The largest amplitude in $\J$ occurs at $t/\tau=0.25$ which is also the time of 
the largest change in the density profile.
The constructed external force, shown in Fig.~\ref{fig3}c), is highly non-trivial
and it is closely related to the behaviour of $\Jdot$, shown in panel Fig.~\ref{fig3}d).
Initially, the external force accelerates the particles towards the maxima
of the one-body density. Then, around $t/\tau=0.25$ the external
force flips sign and decelerates, therefore, the particles.
At $t/\tau=0.5$ there is a jump in the time evolution of the external force due
to the imposed vanishing $\Jdot$ (compare the profiles at
$t/\tau=0.499$ and $0.502$).
Custom flow finds the correct external force despite this drastic change
in time. 
After $\Jdot$ vanishes, both $\rho$ and $\J$ do not
change anymore. Interestingly, the external force continues to evolve in time.
This can be explained by memory effects occurring in both the internal
force field $\fint$, Fig.~\ref{fig3}e), and the transport term $\divtau$, Fig.~\ref{fig3}f).
Even though $\Jdot$ vanishes at $t/\tau>0.5$, the external force 
still needs to vary in time in order to cancel the time evolution
of the internal and the transport terms. Custom flow 
is therefore a valuable tool to study memory effects~\cite{meyer2020non,lesnicki2016molecular,jung2017iterative,berner2018oscillating,treffenstadt2020memory} since
it allows to isolate memory contributions in the force balance equation
from the time evolution of the density and the current fields.

\section{Discussion and conclusions}\label{sec:Conclusion}
We have presented a numerical iterative scheme, see eqs.~\eqref{eq:itegeneral} and \eqref{eq:fextJ}, to construct the external
force required to achieve a given (prescribed) time evolution of a Newtonian
many body-system, as specified by the one-body density and the one-body current.
We have previously shown
how in overdamped Brownian dynamics the exact one-body
force balance equation can be directly used to construct a
reliable custom flow method~\cite{de2019custom}. The external
force is generated as the sum of different contributions that
are sampled in the simulation.
Here, we have followed a different and more general approach.
We construct iteratively the external force by adding at each
iteration terms that correct the external force in the right
direction and that vanish when target and sample fields coincide.
Although we have restricted ourselves here to inertial molecular dynamics,
the method is general and can be also used in e.g. overdamped Brownian
dynamics and Langevin dynamics. There the corresponding force balance
equation can be used to recalculate suitable expressions for the prefactors
$\alpha$, $\beta$, and $\gamma$ since they might be different.

The more general iterative scheme, eq.~\eqref{eq:itegeneral},
modifies the external force based on three types of target-sampled differences that occur in
the gradient of the density, in the current, and in the time derivative of the current.
This is analogous to the three different fundamental equations in classical mechanics: the d'Alembert's principle
based on particle displacements, the Jourdain's principle based on variations of particle velocities,
and the Gibbs-Appell-Gauss's principle based on variations of particle accelerations. It is therefore
not surprising that only one prefactor $\alpha$, $\beta$, or $\gamma$ in eq.~\eqref{eq:itegeneral} need to be present
(using only $\gamma$ is restricted to curl-free target current profiles as discussed in Sec.~\ref{CF}).

Since the calculation of the external force requires to sample only the current,
but not the individual contributions to the force balance equation, it is straightforward
to use custom flow together with a thermostat.
Applying custom flow to systems with the same
target fields but different thermostat algorithms provides new insight into
the different mechanisms to control the temperature since it is possible to precisely analyse
how the individual contributions to the force balance equation are affected by different thermostats.
The choice of thermostat can heavily
influence the results~\cite{wong2010static,thomas2015thermostat} and custom flow might help
to make an educated selection on how to control the temperature out-of-equilibrium,
which is a delicate issue~\cite{nonequilibriumtemperature}.

Custom flow can be used to prescribe target fields such that at least
one contribution to the force balance equation vanishes. For example,
the current vanishes after a certain time in the example of Sec.~\ref{IIID}.
This can facilitate the study of memory effects and the
structure of memory kernels, a topic of current
interest~\cite{meyer2020non,lesnicki2016molecular,jung2017iterative,berner2018oscillating,treffenstadt2020memory}.

The external forces constructed with custom flow are in general noisy
since they are tailored to the finite set of initial microstates used during the iterative process.
Nevertheless, we have shown that a smoothed version of the external force,
constructed by filtering out the high frequency terms, also
produces the target dynamics within the numerical accuracy.
We note however that we have stayed away from instabilities
that might be the source of convergence issues.

Although they are only model situations, the examples considered here are
demanding; the target fields vary substantially over distances comparable
to the particle size and we have designed a case in Sec.~\ref{IIID} for which the resulting external
force is discontinuous in time. Custom flow has in all cases found
the external force field that produces the target fields within the numerical accuracy.
Nevertheless, convergence issues can occur
in e.g.\ strongly driven systems, flows with rapid spatial variations, 
and near the onset of mechanical and fluid instabilities.
Integer arithmetic~\cite{Levesque1993,hoover2020time} and using small values 
for the prefactors $\alpha$, $\beta$, and $\gamma$ might help to mitigate
some of the problems that might appear. 

The existence of a unique mapping between the density distribution and a time-dependent
external potential is at the core of time-dependent density functional theory~\cite{PhysRevLett.94.183001}.
Such mapping is not completely general but restricted to the occurrence of gradient-like forces only.
Similarly, in the widely spread dynamical density functional theory~\cite{Evans1979,Marconi1999},
the internal force field is drastically approximated as a functional of the density distribution only.
No functional dependence on the flow occurs.
These limitations are solved in the formally exact power functional
theory~\cite{PowerF,PFTMD} that considers a functional dependence on all kinematic fields. Such dependence is
required to properly describe e.g. shear migration~\cite{PRLstr}, phase coexistence of active particles~\cite{active},
and laning formation in binary mixtures~\cite{laning}.
Power functional theory relies on a mapping between the external force and both the density and the current distributions. 
Such mapping is indispensable to e.g.\ describe systems in which the current field contains non-gradient
contributions (i.e. rotational and harmonic contributions) since the continuity equation links only the
divergence of the current and the time evolution of the density profile. It is therefore perfectly possible to construct
families of systems that e.g.\ share the same time evolution of the density profile but have
different current profiles~\cite{de2020flow} and are therefore generated by different
external forces. Custom flow provides the numerical evidence of the existence of the unique mapping between
the external force and the kinematic fields.

Custom flow has proven to be an excellent tool to develop 
approximated power functionals in overdamped Brownian systems~\cite{de2020flow} and
we expect it to be also of great help to develop approximate power functionals 
in Newtonian systems. To study large scale systems it can be useful to extend
custom flow to adaptive resolution techniques for multiscale molecular
dynamics simulations~\cite{PhysRevE.67.046704,doi:10.1063/1.2132286,PhysRevLett.110.108301,doi:10.1063/1.5031206}.

\begin{acknowledgments}
We thank Sophie Hermann for a critical reading of the manuscript.
This work is supported by the German Research Foundation (DFG) via
project number 447925252.
\end{acknowledgments}

\appendix
\section{Prefactors from the force balance equation}\label{append}

We start with the exact one-body force
balance equation~\eqref{eq:MDonebodyforcebalance}. Solving for
the external force field yields
\begin{equation}
	\fext(\textbf{r},t)=\frac{m\Jdot(\textbf{r},t)}{\rho(\textbf{r},t)} - \fint(\textbf{r},t) - \frac{\divtau(\textbf{r},t)}{\rho(\textbf{r},t)}.\label{solvefext}
\end{equation}
Following the ideas of Ref.~\cite{de2019custom}, it is possible to establish an iteration scheme
to find the external force at iteration $k+1$ as
\begin{equation}
	\fext^{(k+1)}(\textbf{r},t)=\frac{m\Jdot(\textbf{r},t)}{\rho(\textbf{r},t)} - \fint^{(k)}(\textbf{r},t) - \frac{\divtau^{(k)}(\textbf{r},t)}{\rho(\textbf{r},t)}.\label{eq:iteri+1}
\end{equation}
Here the unknown terms on the right hand side of eq.~\eqref{solvefext}, i.e. the internal
force field and the kinetic stress tensor, are sampled at each iteration and
used to construct the external force iteratively. This idea, that works in
overdamped Brownian systems~\cite{de2019custom},
presents stability issues in inertial systems and does not converge in general.
We next observe that the one-body force balance equation~\eqref{eq:MDonebodyforcebalance} implies that for iteration $k$
\begin{equation}
    \fext^{(k)}(\textbf{r},t)=\frac{m\Jdot^{(k)}(\textbf{r},t)}{\rho^{(k)}(\textbf{r},t)} - \fint^{(k)}(\textbf{r},t) - \frac{\divtau^{(k)}(\textbf{r},t)}{\rho^{(k)}(\textbf{r},t)}.\label{eq:iteri}
\end{equation}
Combining eqs.~\eqref{eq:iteri+1} and \eqref{eq:iteri} yields
\begin{eqnarray}
	\fext^{(k+1)}(\textbf{r},t)&=&\fext^{(k)}(\textbf{r},t)+\frac{m}{\rho(\textbf{r},t)}\left(\Jdot(\textbf{r},t)-\Jdot^{(k)}(\textbf{r},t)\right),\nonumber\\
	&&-\divtau^{(k)}(\textbf{r},t)\left(\frac1{\rho(\textbf{r},t)}-\frac1{\rho^{(k)}(\textbf{r},t)}\right)\label{eq:fextJdot}
\end{eqnarray}
where we have assumed that target and sampled density profiles are the same, i.e. set $\rho^{(k)}(\textbf{r},t) \to \rho(\textbf{r},t)$
in the first term of the right hand side of eq.~\eqref{eq:iteri}. Note that this is necessarily the case if the iterative process converges.
Comparing eqs.~\eqref{eq:fextJdot} and~\eqref{eq:itegeneral} yields $\beta=\frac m{\rho}$.

To find an expression for $\alpha$ 
we approximate $\Jdot(\textbf{r},t)$ and $\Jdot^{(k)}(\textbf{r},t)$ in eq.~\eqref{eq:fextJdot} by
\begin{eqnarray}
	\Jdot(\textbf{r},t)&=&\dfrac{\J(\textbf{r},t)-\J(\textbf{r},t-\Delta t)}{\Delta t},\label{eq:jd1}\\
	\Jdot^{(k)}(\textbf{r},t)&=&\dfrac{\J^{(k)}(\textbf{r},t)-\J(\textbf{r},t-\Delta t)}{\Delta t},\label{eq:jd2}
\end{eqnarray}
where we have used that the sampled $\J^{(k)}$ and the target $\J$
coincide at time $t-\Delta t$. This is again necessarily the case
if the process converges. Although a central time difference would be more precise,
we use here the backward time difference since $\J^{(k)}(\textbf{r},t+\Delta t)$ is
unknown at time $t$.
Inserting eqs.~\eqref{eq:jd1} and ~\eqref{eq:jd2} into eq.~\eqref{eq:fextJdot} and
comparing the result to eq.~\eqref{eq:itegeneral} results in $\alpha=\frac m {\rho\Delta t}$.

Finally to find a suitable expression for $\gamma$ we use the equilibrium expression for the transport
term~\eqref{eq:divtau} to roughly approximate the second term in eq.~\eqref{eq:fextJdot} by
\begin{equation}
	-\divtau^{(k)}\left(\frac1{\rho}-\frac1{\rho^{(k)}}\right)\sim+k_BT\nabla\ln\frac{\rho}{\rho^{(k)}}.~\label{eq:blabla}
\end{equation}
From which we obtain $\gamma=k_BT$ by comparison to eq.~\eqref{eq:itegeneral}. We note that
eq~\eqref{eq:blabla} is a crude approximation but we are only interested in a suitable expression
for the prefactor $\gamma$. The precise values of the prefactors is not critical
for the method to converge (provided
they are small enough such that the iterative process is stable). However, having suitable expressions
is relevant to achieve a fast convergence since the prefactors
control the amount of change of the external force from iteration to iteration.

\normalem

%

\end{document}